\documentclass[10pt,a4paper]{article}

\usepackage{settings}
\usepackage{code}

\usepackage{subcaption}
\usepackage{hyperref}

\usepackage{tikz,pgfplots,pgfplotstable}
\usetikzlibrary{patterns}

\newcommand{\memPlot}[4]{
  \addplot+[#3] table[y=#1, meta=Label, x expr=\coordindex] {#4};
  \addlegendentry{#2}
}

\pgfplotsset{ %
  cycle list={ %
    blue, mark=square*\\%
    red, mark=*\\%
    orange, mark=triangle*\\%
    black!40!green, mark=diamond*\\%
  },
}

\pgfplotsset{ %
  cycle list={ %
    blue, mark=square*\\%
    red, mark=*\\%
    orange, mark=triangle*\\%
    black!40!green, mark=diamond*\\%
  },
}

\DeclareMathOperator{\proj}{proj}

\renewcommand{\Re}{\text{Re}}
\renewcommand{\Im}{\text{Im}}

\title{Adding complex numbers to expression template algorithmic differentiation tools}
\author{Max Sagebaum\footnote{Contact: max.sagebaum@scicomp.uni-kl.de}, Nicolas R. Gauger}

\date{
  Chair for Scientific Computing,\\ University of Kaiserslautern-Landau (RPTU), Germany\\[2ex]
  August 7, 2025
}

\begin{document}

\maketitle
\begin{abstract}
Operator overloading algorithmic differentiation (AD) tools are usually only developed for floating-point values. Algorithmic optimization for, e.g., linear systems solvers or matrix-matrix multiplications are often introduced via external functions or manual function specializations. Complex numbers can be viewed as aggregates of two floating-point values on which specialized operations are applied. Typically, these operations can be handled by the regular floating-point operations from the AD tool. Nevertheless, adding the complex number operations to the expression template framework of modern operator overloading AD tools has several benefits. The internal computations of a complex number operation are hidden, and the complex operations do not decompose into single operations. This leads to a smaller memory footprint of the recorded tape and faster gradient computation times. We will discuss these problems, analyze how complex numbers can be integrated into modern operator overloading AD tools, demonstrate an implementation in CoDiPack, and show performance results on a synthetic test case.
\end{abstract}

\section{Introduction}

The application of algorithmic differentiation (AD) to a computer program enables this program to evaluate the derivatives alongside the regular (primal) computation. With the operator overloading approach, the computation type in the program is exchanged with the so-called \emph{active type} of the AD tool. The reverse mode (back propagation) stores data like internal identifiers and Jacobian entries during the regular computation for each operation (e.g., *, +, sin, cos) on a tape (stack). Afterward, the data on the tape is interpreted in the reverse order for the computation of the derivatives.

The set of operations for the active type usually covers all the basic operations used in a computer program. Algorithms that are formed with these operations are recorded on the tape, and the derivatives can be computed. The problem with this approach is that it is not optimal for all algorithms. This is demonstrated in \cite{eigenAD} for the matrix-matrix product multiplication and the solution of linear systems in Eigen. The set of operations is extended for the active type in dco/c++ \cite{AIB-2016-08} such that the matrix-matrix product and solution of a linear system are now known by dco/c++. For both operations, this resulted in a reduced amount of memory that is written to the tape and an improved runtime. In the case of the linear system solve, the runtime improves from a complexity of $n^3$ to $n^2$. In the same way, the AD tool Adept \cite{hogan2014fast} has been specialized for array and vector operations such that these are stored on the tape in an optimized way. The Stan math library \cite{carpenter2015stan} provides optimized routines for taping linear algebra operations as well.

% tanh: stmt: 6 jac: 9 tape: 5 * 6 + 12 * 9 = 138
% tanh opt: stmt: 2 jac: 4 tape: 5 * 2 + 12 * 4 = 58

As it has been done in the papers above, the operations for complex number types can also be optimized from an AD viewpoint. Since complex numbers are just an aggregate of two active types, each operation on a complex number performs several steps to compute the result. An example is the complex \ic{tanh} function in the stdc++ library\footnote{\url{https://libcxx.llvm.org}}. This function uses a total of 9 operations on active types, which requires 138 bytes of tape storage for a Jacobian taping approach \cite{SaAlGauTOMS2019}. A specialized handling of the complex \ic{tanh} would use tape storage of 58 bytes, which is a memory reduction of 55 \%.

The expression template technique \cite{hogan2014fast, SaAlGauTOMS2019} allows multiple nested operations to be recorded as one large operation by AD. The equation
\begin{equation}
w = \sqrt{u^2 + v^2}
\label{eq.ooExample}
\end{equation}
is recorded as one operation with two arguments and not as four separate operations as shown in Listing \ref{lst.ooExample}.

Without any specialized handling for complex numbers, the statement in Equation \eqref{eq.ooExample} would again break down into the four intermediate operations. The tape storage for these four operations with complex numbers would be 232 bytes, assuming that all operations are already optimized for AD. If the statement in Equation \eqref{eq.ooExample} does not separate into intermediate operations and can be captured by the AD tool as one large expression, then the tape storage would be 106 bytes, which is a reduction by about 50 \% for the complex number case.

% separated stmt: 8 jac: 16 tape: 5 * 8 + 16 * 12 = 232
% optimized stmt: 2 jac: 8 tape: 5 * 2 + 8 * 12 = 106
\begin{codeRef}[t]{lst.ooExample}{Separation of the statement $w = \sqrt{u^2 + v^2}$ into intermediate operations.}
  t1 = pow(u, 2);
  t2 = pow(v, 2);
  t3 = t1 + t2;
  w = sqrt(t3);
\end{codeRef}

The demonstrated memory reductions are just for one example operation and statement. The 55 \% reduction for \ic{tanh} will not apply to all operations, since only a few complex operations require several intermediate steps for the computation. But the expression template handling of statements will always reduce the required tape memory. The total amount of the reduced memory will depend on the statements in the code, but as a rule of thumb, a reduction of about 40 \% can be expected. For example, a tape that required 1 GB of memory for the complex operations will be reduced to 600 MB of memory with the specialized handling.

In this paper, we will demonstrate how operations for complex numbers can be specialized for AD and how they can be integrated into the expression template framework. In addition, we will explore how the extent expressions can be stored in a Jacobian taping approach and a primal value taping approach \cite{SaAlGa2018OMS}.
The presented work is structured as follows: Section \ref{sec.ADIntroduction} will give a brief introduction to AD and expression templates. Afterward, Section \ref{sec.expressionTemplateExtension} will extend this theory for complex numbers. Section \ref{sec.tapeDataLayout} will discuss how the extended expressions can be stored on the common taping strategies for AD, which is followed in Section \ref{sec.implementation} by a discussion of the implementation and pitfalls during development. Finally, Section \ref{sec.results} will present performance measurements for a generic test case. All code in this paper is C++ code.

\section{AD theory and expression templates}
\label{sec.ADIntroduction}

In this section, we give a brief introduction to AD. For a complete overview, please see \cite{grie08, naumann2012art}.

We assume that each program can be described as the function \ic[breaklines]{void func( double const x[], double y[])} where \ic{x} are the input values and \ic{y} are the output values. How \ic{func} is programmed and if it even exists is not relevant, since it only describes the general setup for the theory. We can then define the mathematical function $F: \R^n \rightarrow \R^m$ with $y = F(x)$. Since $F$ is defined by the implementation \ic{func}, we can always assume that $F$ can be broken down into several elementary functions 
\begin{equation}
\phi_i: \R^{d_i} \rightarrow \R
\label{eq.elementalFunction}
\end{equation}
with $w_i = \phi(v_i)$ and $i = 1 \ldots N$. Elementary functions can be operations like $*$, $+$ or functions like $\sin$, $\cos$. The range of $N$ is arbitrary, but for industrial applications, the number of elementary functions evaluated can be in the region of $10^{12}$.

The evaluation of $F$ is carried out by first evaluating $\phi_1$ then $\phi_2$ until $\phi_N$ is reached. Each elemental function is based on intermediate values that are computed by previous elemental functions. During this process, the input variables $x$ are read and the output variables $y$ are written. That is, $F$ can be interpreted as a concatenation of all $\phi_i$. By applying the chain rule and the directional derivative to $F$ and therefore the concatenation of the $\phi_i$ the \emph{forward mode} of AD is defined as
\begin{equation}
  \dot y = \frac{d F}{d x}(x) \dot x
  \label{eq.forwardMode}
\end{equation}
where $\dot x \in \R^n$ defines the direction for the forward derivative and $\dot y \in \R^m$ is the directional derivative of $F$ in the direction $\dot x$. AD does not build the Jacobian $\frac{d F}{d x}$ directly, the result $\dot y$ is computed by evaluating the directional derivative for each elemental function $\phi_i$ alongside the primal evaluation
\begin{equation}
  \dot w_i = \frac{d \phi_i}{d v_i}(v_i) \dot v_i \quad \forall i = 1 \ldots N \eqdot
  \label{eq.forwardModeElemental}
\end{equation}

The \emph{reverse mode} of AD is derived by building the discrete adjoint of the forward mode equation \eqref{eq.forwardMode}. The AD reverse mode computes
\begin{equation}
  \bar x = \frac{d F}{d x}^T(x) \bar y
\end{equation}
where $\bar y$ is the seeding for the reverse propagation of the derivatives and $\bar x$ is the resulting derivative. $\bar x$ is spoken as \emph{x bar} and is not the conjugate complex operator for complex numbers. The bar notation is the standard notation in reverse AD for the adjoint variables. The conjugate complex is denoted via the complex transposed operator ($\circ^H$) in this paper.

As in the forward mode, the Jacobian $\frac{d F}{d x}^T$ is not build directly, it is computed by the reversal of all elemental functions. That is, for each elemental function, the adjoint equation
\begin{equation}
  \bar v_i \aeq \frac{d \phi_i}{d v_i}^T(v_i) \bar w_i \quad \forall i = N \ldots 1
  \label{eq.reverseModeElemental}
\end{equation}
is evaluated in reverse order from $N$ to $1$.

Because the elemental functions can no longer be evaluated alongside the primal evaluation in the reverse mode of AD, the AD implementations break down the reverse mode into two phases. The first phase is the recording phase, which is enabled during the primal computation. Here, the AD tool stores all the information required to evaluate Equation \eqref{eq.reverseModeElemental}. The information is stored in a structure called tape, which usually behaves like a stack. In the second phase, the tape is interpreted from the last element to the first element, and for all elemental functions, Equation \eqref{eq.reverseModeElemental} is evaluated.

AD can be applied to a code through source transformation \cite{Hascoet2013TTA} or operator overloading \cite{AIB-2016-08, SaAlGauTOMS2019}. Source transformation AD parses the source code of the program and generates a new source code that is augmented by either the forward or reverse AD operations. In operator overloading AD, the AD tool provides a new computation type that is typically called the active type. This type is used instead of the regular computation type like \ic{double} in the application, which allows the AD tool to see all the executed operations and perform the necessary forward or reverse operations.

In this paper, we are focusing on the operator overloading approach.

\subsection{Expression templates}
The expression template technique was first introduced by Aubert \cite{Aubert2001} and is used in several AD tools \cite{SaAlGauTOMS2019, hogan2014fast, AIB-2016-08}. We just give here a brief introduction, please see \cite{hogan2014fast, SaAlGauTOMS2019} for a more in-depth discussion.

Operator overloading AD tools define a set of operations that can be handled by the active type \ic{AReal}. Usually, these are binary operations \ic[breaklines]{AReal op(AReal, AReal)} like \ic{+}, \ic{*} or \ic{pow} or unary operations \ic{AReal op(AReal)} like \ic{-}, \ic{sin} or \ic{tanh}. Statements in the code like
\begin{codeRefInline}{lst.fullExampleExpression}{Example statement.}
w = sqrt(pow(u, 2) + pow(v, 2));
\end{codeRefInline}
are separated into intermediate operations as shown in Listing \ref{lst.ooExample}. The AD tool sees four operations that are handled one by one.

Expression templates \cite{Veldhuizen1995} in combination with a lazy evaluation \cite{wadsworth1971semantics,Henderson:1976:LE:800168.811543} allow changing the patterns of the operations. For binary operations, this is then \ic{Op<A, B> op(A a, B b)} where \ic{A} and \ic{B} are template arguments and \ic{Op} is the implementation of the operation logic. That \ic{Op} is based on the arguments employs the expression template technique, and that the operation returns an object that can evaluate the operation employs the lazy evaluation technique. The advantage of this change in the definition of the operators is that the statement in Listing \ref{lst.fullExampleExpression} creates the operation
\begin{code}
  Sqrt<Add<Pow<AReal, Constant>, Pow<AReal, Constant>>>
\end{code}
for the right-hand side of the assignment. In the assignment, the AD tool can now examine all operations on the right-hand side and can optimize the stored data for the assignment. In this case, it stores one statement with two arguments instead of four operations with a total of five arguments.

\section{Expression templates for complex numbers}
\label{sec.expressionTemplateExtension}

The AD theory described in Section \ref{sec.ADIntroduction} assumes that each elemental function as defined in Equation \eqref{eq.elementalFunction} has only one output argument. Since complex numbers are composed of two floating-point values, complex operations are not covered by this definition. As done by Griewank and Walter in \cite{grie08}, the theory of AD can be extended such that elemental functions with an arbitrary number of output values are allowed. The definition of the elemental functions is then
\begin{equation}
\phi_i: \R^{d_i} \rightarrow \R^{p_i} \eqdot
\label{eq.elementalFunctionExtended}
\end{equation}
It is now possible that an operation has multiple output values, which has a significant impact on the data layout for the primal value taping strategy, which we will discuss in Section \ref{sec.tapeDataLayout}. In this section, we continue with the extension of the expression templates for complex numbers.

The operations defined for \ic{std::complex} are
\begin{itemize}
  \item Binary operations: \ic{operator +}, \ic{operator -}, \ic{operator *}, \ic{operator /}, \ic{pow}
  \item Binary real valued operations: \ic{polar}
  \item Unary operations: \ic{operator +}, \ic{operator -}, \ic{real}, \ic{imag}, \ic{abs}, \ic{arg}, \ic{norm}, \ic{conj}, \ic{proj}, \ic{exp}, \ic{log}, \ic{log10}, \ic{sqrt}, \ic{sin}, \ic{cos}, \ic{tan}, \ic{asin}, \ic{acos}, \ic{atan}, \ic{sinh}, \ic{cosh}, \ic{tanh}, \ic{asinh}, \ic{acosh}, \ic{atanh}.
\end{itemize}
For all these operations, the derivative definitions can be found in \cite{olver2010nist} or worked out by hand. For the multiplication $M: \C \times \C \rightarrow \C$ with $w = M(a, b) = a * b$ the derivatives are
$\frac{d M(a,b)}{d (a,b)} = (b, a) \in \C^2$. For the forward mode of AD, the elemental function update from Equation \ref{eq.forwardModeElemental} becomes
\begin{equation}
\dot w = \frac{d M(a,b)}{d (a,b)} \colvec{2}{\dot a}{\dot b} = b \dot a + a \dot b
\end{equation}
which is equivalent to the real valued implementation. For the reverse mode of AD, the elemental function update from Equation \ref{eq.reverseModeElemental} becomes
\begin{equation}
\bar v_i \aeq \frac{d M(a,b)}{d (a,b)}^H \bar w = \colvec{2}{b^H \bar w}{a^H \bar w} \eqdot
\end{equation}
Please note that the transposed in Equation \ref{eq.reverseModeElemental} becomes the complex transposed, denoted by $\circ^H$ in the complex setting. Therefore, in the complex case, the implementation reverse mode differs from the real valued implementation.

All binary operations have, in addition to the variants for complex numbers, the variants with mixed types. In total, there are three overloads for each binary operation that have the arguments \ic{(complex<T>, complex<T>)}, \ic{(complex<T>, T)} and \ic{(T, complex<T>)}, which all need to be considered. If we take a look at the complex multiplication with one scalar argument $M_s: \C \times \R \rightarrow \C$ with $w = M_s(a, \beta) = a * \beta$ then the derivatives are defined as $\frac{d M_s(a,\beta)}{d (a,\beta)} = (\beta, a)$ which looks the same as for $M$. The forward mode AD elemental function update looks also the same as for $M$
\begin{equation}
\dot w = \frac{d M_s(a,\beta)}{d (a,\beta)} \colvec{2}{\dot a}{\dot \beta} = \beta \dot a + a \dot \beta \eqdot
\end{equation}
If we write now the reverse mode AD element function update
\begin{equation}
\bar v_i \aeq \frac{d M_s(a,\beta)}{d (a,\beta)}^H \bar w = \colvec{2}{\beta \bar w}{a^H \bar w}
\label{eq.complexMulReverseWrong}
\end{equation}
then we see that the update is wrong since $\bar v_i \in \C \times \R$ and the right-hand side of Equation \eqref{eq.complexMulReverseWrong} is an element of $\C \times \C$. The proper solution to this discrepancy can be derived when we interpret $\C$ as $\R^2$ and take a look at the function $C: \R \rightarrow \C$ with $w = C(\alpha) = (\alpha, 0)^T$. The derivative of $C$ is $\frac{d C}{d \alpha} = (1, 0)^T$ with $\frac{d C}{d \alpha}^H = (1, 0)$. We use the complex transposed since we are just interpreting $\C$ as $\R^2$. The reverse mode AD elemental function evaluation of $C$ is then 
\begin{equation}
\bar \alpha \aeq \frac{d C(\alpha)}{d \alpha}^H \colvec{2}{\Re(\bar w)}{\Im(\bar w)} = (1, 0)\colvec{2}{\Re(\bar w)}{\Im(\bar w)} = \Re(\bar w)
\end{equation}
where $\Re(\circ)$ is the real part of a complex number and $\Im(\circ)$ the imaginary part. If we define $M_s(a, \beta) := M(a, C(\beta))$ and apply the reverse mode of AD to this definition, then Equation \ref{eq.complexMulReverseWrong} becomes 
\begin{equation}
\bar v_i \aeq \frac{d M_s(a,\beta)}{d (a,\beta)}^H \bar w = \colvec{2}{\beta \bar w}{\Re(a^H \bar w)} \eqdot
\label{eq.complexMulReverseCorrect}
\end{equation}
Without loss of generality, the same strategy can be applied for the other mixed-value binary operations. The result is that for real valued arguments, only the real part of the computed reverse mode AD update is applied.

In terms of the implementation, $C$ can be thought of as a constructor, and $\frac{d C}{d \alpha}^T$ is the discrete adjoint of the constructor.

\section{Tape data layout}
\label{sec.tapeDataLayout}

In this section, we will analyze the data layout for a Jacobian taping strategy \cite{SaAlGauTOMS2019} and a primal value taping strategy \cite{SaAlGa2018OMS}.

\subsection{Jacobian taping}

The Jacobian taping strategy described in \cite{SaAlGa2018OMS} which is also used by dco/c++ \cite{AIB-2016-08} and adept \cite{hogan2014fast} computes the Jacobian of the elemental function from Equation \eqref{eq.elementalFunction} and stores it on the tape. The Jacobian is used to evaluate the reverse mode AD Equation \eqref{eq.reverseModeElemental} for the elemental function. Since the number of inputs $d_i$ is small, the size of the Jacobian per element function is manageable. As described in \cite{SaAlGauTOMS2019} the adjoint (e.g., $\bar a$) of a primal variable (e.g., $a$) is identified with an index, that is, a lookup into the adjoint vector. Therefore, the storage for an elemental operation is
\begin{itemize}
  \item 1 byte for $d_i$,
  \item 4 byte for the identifier of $\bar w$,
  \item $8 \cdot d_i$ bytes for the Jacobian $\frac{d \phi_i}{d v_i}$ and
  \item $4 \cdot d_i$ bytes for the identifiers of $\bar v_i \in \R^{d_i}$.
\end{itemize}

In the setting of complex numbers, we are no longer storing elemental functions as described in Equation \eqref{eq.elementalFunction}, but instead look at the extended version described in Equation \eqref{eq.elementalFunctionExtended}. In the pure complex settings, the number of outputs $p_i$ is 2 and the number of inputs $d_i$ is a multiple of 2, since every complex number consists of two values. We have to store now the full Jacobian matrix $\frac{d \phi_i}{d v_i} \in \R^{p_i \times d_i}$ and all identifiers for input adjoints $\bar v_i$ and output adjoints $\bar w$. This would mean that the new data layout of the tape would look like
\begin{itemize}
  \item 2 bytes for $d_i$ and $p_i$,
  \item $4 \cdot p_i$ bytes for the identifiers of $\bar w$,
  \item $8 \cdot d_i \cdot p_i$ bytes for the Jacobian $\frac{d \phi_i}{d v_i}$ and
  \item $4 \cdot d_i$ bytes for the identifiers of $\bar v_i \in \R^{d_i}$.
\end{itemize}
This data layout has the problem, that elemental functions with only one output still need to store $p_i = 1$. In a worst-case scenario with $d_i = 1$, this is a 6 \% increase in tape size. Another disadvantage is, that always the full Jacobian $\frac{d \phi_i}{d v_i}$ needs to be stored. In the case where $p_i$ is always one, that is without the extension for complex numbers, zero entries of the Jacobian do not need to be recorded, which saves 12 bytes per zero entry. This memory optimization is described in more detail in \cite{SaAlGa2018OMS}. With $p_i \not = 1$, this technique can only be applied when the whole column for the input is zero. One example where this is quite relevant is the complex addition or subtraction. For the addition operation the Jacobian is $\colvec{2}{1 0 1 0}{0 1 0 1}$ which would require in the above setting $90$ bytes where $32$ bytes are used to store zeros.

Because of these two reasons, memory increase in the tape for the case $p_i = 1$ and zero Jacobian entries cannot be dropped, we decided to keep the memory layout of the Jacobian tape and store each assignment of Equation \eqref{eq.elementalFunctionExtended} as a separate statement.

As always, self references in the elemental functions need to be considered. Let us consider \ic{c *= a} as an example. This is the mathematical equivalent to 
\begin{equation}
c_n = c_c \cdot a \ \ ,
\label{eq.exampleSelfReference}
\end{equation} where $c_c \in \C$ is the current value of $c$ and $c_n \in \C$ is the new value. With the proposed strategy, we would first store $\Re(c_n) = \Re(c_c \cdot a)$ and afterward $\Im(c_n) = \Im(c_c \cdot a)$. Since $c_n$ and $c_c$ have the same memory position, the real part of $c_c$ is already changed when the first equation is stored. Therefore, the result of $\Im(c_c * a)$ is changed. This can be overcome by first storing all data for the right-hand side for both equations. Afterward, the data for the left-hand side can be updated and stored. 

The reverse evaluation of Equation \eqref{eq.exampleSelfReference}
\begin{subequations}
  \begin{align}
    \Re(\bar c_o) & \aeq \Im(a) \cdot \Im(\bar c_n); & \Im(\bar c_o) & \aeq \Re(a) \cdot \Im(\bar c_n);
      \label{eq.examplesSelfReferenceReverse_A} \\
    \Re(\bar a) & \aeq \Im(c_o) \cdot \Im(\bar c_n); & \Im(\bar a) & \aeq \Re(c_o) \cdot \Im(\bar c_n); \\
    \Im(\bar c_n) & = 0 &&\\
    \Re(\bar c_o) & \aeq \Re(a) \cdot \Re(\bar c_n); & \Im(\bar c_o) & \aeq -\Im(a) \cdot \Re(\bar c_n); \label{eq.examplesSelfReferenceReverse_D} \\
    \Re(\bar a) & \aeq \Re(c_o) \cdot \Re(\bar c_n); & \Im(\bar a) & \aeq -\Im(c_o) \cdot \Re(\bar c_n); \\
    \Re(\bar c_n) & = 0 &&
  \end{align}
  \label{eq.examplesSelfReferenceReverse}
\end{subequations}
consists of two separate reverse evaluations since we decided to store Equation \ref{eq.exampleSelfReference} in such a way. The same problem as in the storing can occur when $\bar c_n$ and $\bar c_c$ have the same position in the adjoint vector. $\Re(\bar c_o)$ is updated in Equation \eqref{eq.examplesSelfReferenceReverse_A}, which is then used in Equation \ref{eq.examplesSelfReferenceReverse_D} as $\Re(\bar c_n)$ again and leads to wrong results. 

The solution to this issue is a modification of the storing procedure. The indices of the left hand side values are not reused for aggregated types. New one are acquired before the old left hand side indices are freed.

\subsection{Primal value taping}
\label{sec.dataLauoutPrimalValueTaping}

The primal value taping strategy described in \cite{SaAlGa2018OMS} stores the primal values of the computation and stores function pointers for the reverse mode AD computation for each elemental function (Equation \eqref{eq.reverseModeElemental}). The Jacobian is computed on the fly during the reverse interpretation of the tape. As in the Jacobian taping approach, identifiers for the bar values are used. Therefore, the storage for an elemental operation is
\begin{itemize}
  \item 1 byte for the number of inactive arguments $n_i^\mathrm{inactive}$ (identifier of the adjoint is zero),
  \item 8 bytes for the function pointer to the reverse evaluation of Equation \ref{eq.reverseModeElemental},
  \item 4 bytes for the identifier of $\bar w$,
  \item 8 bytes for the primal value of $w$,
  \item $4 \cdot d_i$ bytes for the identifiers of $\bar v_i \in \R^{d_i}$,
  \item $n_i^\text{inactive} \cdot 8$ bytes for the primal values of inactive arguments and
  \item $n_i^\text{constant} \cdot 8$ bytes for the primal values of constant arguments.
\end{itemize}
$n_i^\text{constant}$ describes the count of constant arguments like $4.0$ in $w = 4.0 \cdot a$. It does not need to be stored on the tape since it is available in the function pointer for the reverse evaluation. The same is true for the number of input arguments $d_i$, which is therefore not stored on the tape.

As for the Jacobian taping approach, we are switching now to the extended elemental function Equation \eqref{eq.elementalFunctionExtended}, where $p_i$ is equal to 2 and $d_i$ is a multiple of 2 in a pure complex setting. As for $d_i$, the number of outputs $p_i$ is indirectly stored in the function pointer and does not need to be recorded on the tape. The above tape storage changes only for the third and fourth entry, since they are related to the output values. The new entries are 
\begin{itemize}
  \item $4 \cdot p_i$ bytes for the identifier of $\bar w$ and
  \item $8 \cdot p_i$ bytes for the primal value of $w$.
\end{itemize}
The advantage of the primal value taping approach is, that for the case $p_i = 1$, no additional data is stored. The existing implementation stores the data in four different stacks, the first four entries are stored in the first stack and the last three entries have their own stacks. Since the third and fourth entry have now one item per output, they need to be stored in their own stack which is only a minor change to the implementation.

Unfortunately, the additional number of arguments to the function pointer decreased the performance of the reverse evaluation by about 10\%. The arguments to the function pointers could no longer be stored in the registers according to the x86-64 ABI \cite{x86ABI}. The generated push and pop assembler instructions cause the overhead. This observation lead to a reformulation of the primal value tape implementation. The existing five stacks with specific type information have been discarded and replaced with two stacks. The first stack stores the first two items in the data layout and two additional bytes, that is $n_i^\text{inactive}$, the function pointer, and the total size of the remaining data. The first two items are the ones that are mandatory for each statement. The third item could be avoided, but would make the implementation much more involved and reduce the performance. The second stack stores the remaining data items in a byte data stream.

The data layout of the second stack is from a software engineering standpoint not optimal. Data entries of different sizes are stored, which changes the alignment of the data for each statement. Nevertheless, the performance results in Section \ref{sec.results} show that the new data layout does not affect the reverse interpretation very much. The recording time of primal value tapes improved by approximately $15\ \%$.

The drawback is an overall size increase in primal value tapes by around $5\ \%$.

\section{Implementation}
\label{sec.implementation}

The implementation of the complex type handling in CoDiPack\footnote{\url{https://www.scicomp.uni-kl.de/software/codi/}} is done in a more general fashion. The implementation assumes that an element $d \in D$ of an arbitrary type $D$ can be interpreted as an element $r \in \R^n$ with $r = \proj(d)$, where the projection operator $\proj: D \rightarrow \R^n$ is usually the identity. $d$ is considered an aggregated entity that is defined by a vector of real values. The implementation in CoDiPack implements the type \ic{AggregatedActiveType} to represent such types in the expression framework. This type can be used to implement the specific handling of aggregated types such as complex numbers. A brief description of the basic ingredients in the implementation follows.

The \emph{\textbf{\ic{AggregatedTypeTraits}}} class defines the necessary operations on the aggregated types that are required by CoDiPack. The first operation is the vector access operator $\cdot[\cdot]: D \subseteq \R^n \times \N \rightarrow \R$ with $ w = d[i] := d_i$. The second operation is the array construction $C: \R^n \rightarrow D \subseteq \R^n$ with $d = C(v_1, \ldots, v_n)$. Both operations require the reverse mode AD operation based on Equation \ref{eq.reverseModeElemental}. That is $\bar d \aeq \frac{d \cdot[\cdot]}{d d}^T \bar w$ for the array access and $\bar v_i \aeq \frac{d C}{d v_i}^T \bar d$ for the array construction. Since all four of these implementations are usually quite simple, a base class \ic{AggregatedTypeTraitsBase} exists which implements these operations for the identity embedding.

The \emph{\textbf{\ic{AggregatedActiveType}}} class uses the defined functionality in the \ic[breaklines]{Ag- gregatedTypeTraits} to implement the common operations for all aggregated types. This is the copy constructor and the copy assignment. In addition, it declares the array for the data storage and implements the interface such that the aggregated type fits into the template expression framework of CoDiPack.

The \emph{\textbf{\ic{std::complex<ActiveType>}}} specialization of the \ic{std::complex} type extends from \ic{AggregatedActiveType} and therefore adds the complex numbers of the standard C++ library to the template expression framework of CoDiPack. In addition, the specialization defines all operations on the complex numbers that are not covered by the constructors and assign operators in \ic{AggregatedActiveType}. For the complex numbers these are the construction from a real value, the construction from two real values as well as the assignment operators \ic{+=}, \ic{-=}, \ic{*=} and \ic{/=}.

Certain compilers (e.g., apple clang) have problems with the specialization of \ic{std::complex}. The specialization of \ic{std::complex} can be disabled with the compiler flag \ic{-DCODI_SpecializeStdComplex=0}. A complex active type can then be defined as \ic{codi::ActiveComplex<Real>} where \ic{Real} is a CoDiPack type like \ic{codi::RealReverse}.

\subsection{Member method injection into expression templates}

The complex numbers in C++ have two member operations, \ic{real} and \ic{imag}. Statements with complex numbers can therefore look like
\begin{code}
double w1 = a.real();
double w2 = (a * b).real();
\end{code}
where \ic{a} and \ic{b} are complex numbers. The assignment of \ic{w1} can be handled by adding the \ic{real()} member function to the specialization \ic[breaklines]{std::complex <ActiveType>}. The issue in the \ic{w2} assignment is that \ic{(a * b)} returns an expression and is of the type \ic{ComputeExpression} which does not know about the member operation. It is not an option to extend \ic{ComputeExpression} by the complex member operations since the expression are not only defined for complex numbers and would make no sense for other types, e.g., double values. Since, \ic{ComputeExpression} is also an internal CoDiPack type, users of CoDiPack would not be able to add member operations for their own aggregated types.

In CoDiPack all expressions inherit from \ic{ExpressionInterface}. By the same arguments above, it makes no sense to directly add the complex number member operations to this interface. Instead, we allow the injection of methods into the \ic{ExpressionInterface} by inheritance from a new interface \ic{ExpressionMemberOperations}. The pseudocode for this relation can be seen in Listing \ref{lst.expressionInject}. The template argument \ic{Impl} represents the expression template pattern, and \ic{Real} declares the return type of the expression. If \ic[breaklines]{ExpressionMember- Operations} is now specialized for certain \ic{Real} types like \ic{std::complex<double>}, then all member operations defined in the specialization will be available in all expressions, that inherit from \ic{ExpressionInterface} and have the specialized \ic{Real} type as a return value. The specialization for complex numbers is shown in Listing \ref{lst.expressionInjectComplex}, where the \ic{real()} and \ic{imag()} operations are added to the expressions. The \ic{cast()} method is a helper method that casts the object to the implementation type defined by the template argument \ic{Impl}. Please note that this will also add the member operations to \ic{std::complex<ActiveType>} since it also extends from \ic{ExpressionInterface} via \ic{AggregatedActiveType}. Therefore, the member methods need to be defined only once.

\begin{listing}
  \caption{Injecting member operations into expression templates.}
  \label{lst.expressionInject}
  \begin{code}
// Empty injection class.
template<typename Real, typename Impl>
struct ExpressionMemberOperations {};

// Add operations for all expression types.
template<typename Real, typename Impl>
struct ExpressionInterface : 
    public ExpressionMemberOperations<Real, Impl> { ... };
\end{code}
\end{listing}

\begin{listing}
  \caption{Injecting the complex member operations \ic{real()} and \ic{imag()} into expression templates.}
  \label{lst.expressionInjectComplex}
  \begin{code}
// Specialization for std::complex
template<typename InnerReal, typename Impl>
struct ExpressionMemberOperations<std::complex<InnerReal>, Impl> {
  auto real() {
    return std::real(this->cast());
  }
  
  auto imag() {
    return std::imag(this->cast());
  }
};
\end{code}
\end{listing}

\subsection{Performance issues}
During the implementation of the complex numbers in CoDiPack three performance issues were discovered, that impacted the general performance of CoDiPack, although these changes should not have such a large influence. That the number of arguments to a function pointer call affected the performance is already described in Section \ref{sec.dataLauoutPrimalValueTaping}. The solution is to change the number of stacks and therefore the number of arguments, which circumvents this performance issue.

The next issues are missing inline declarations on default declared constructors and destructors. Since, the generated elemental function operators for the complex numbers are more involved than the real valued counterparts, the heuristics of the compiler decided to not inline these constructors any longer. Adding the missing declarations resolved this issue.

The third issue is also an inline problem. The \ic{std::tuple} constructor is not inlined by the heuristics of the compilers for larger statements with complex numbers. The tuple is used in the generalized implementation of the operator expressions. Since we could not change the standard library implementation, a minimal tuple implementation is added to CoDiPack which can be inlined. The reverse evaluation of the primal values tapes had a slowdown by a factor of $2.5$ because of this issue.

\section{Performance results}
\label{sec.results}

\subsection{Coupled Burgers' equation}
The coupled Burgers' equation is an established test case for the performance comparison of CoDiPack implementations and is described in \cite{SaAlGauTOMS2019} and \cite{SaAlGa2018OMS}. For simplicity, we want to use the same test case for the performance evaluations in this paper. Since the original test case is only formulated for real valued numbers, a few changes for complex numbers have to be made. One consequence is, that the obtained solution is not exact. Nevertheless, we accept this error since we are only interested in performance values and memory consumption.
For completeness, we recapitulate the problem formulation here.

The coupled Burgers' equation \cite{biazar2009exact,bahadir2003fully,zhu2010numerical}
\begin{align}
  u_t + uu_x + vu_y &= \frac{1}{R}(u_{xx} + u_{yy}), \\
  v_t + uv_x + vv_y &= \frac{1}{R}(v_{xx} + v_{yy})
\end{align}
is discretized with a central finite difference scheme since we are using complex numbers.
The initial and boundary conditions are taken from the exact solution given in \cite{biazar2009exact} and are shifted into the complex plane by adding $i$. The modified boundary conditions and initial solution are
\begin{align}
  u(x, y, t) &= \frac{x + y - 2xt}{1 - 2t^2} + i \quad (x,y,t) \in D \times \R,\\
  v(x, y, t) &= \frac{x - y - 2yt}{1 - 2t^2} + i \quad (x,y,t) \in D \times \R \eqdot
\end{align}
The computational domain $D$ is the unit square $D = [0,1] \times [0,1] \subset \R \times \R$.
As far as the differentiation is concerned, we choose the initial solution of the time stepping scheme as input parameters, and as output parameter we take the norm of the final solution.

The node for the test case consists of two AMD EPYC 7262 CPUs with a total of 16 cores and 256 GB of main memory.
We discretize the Burgers' equation on a $601\times 601$ grid and solve it using 16 iterations.
gcc version 15.1 is used as the compiler. We remark that similar results are obtained with the clang compiler as well as on nodes with Skylake CPUs.
All timing values are averaged over 20 evaluations.

There are two load layouts which we use. The \emph{single} load layout runs only one process on the 16 cores, which provides better benchmark results for implementation details, since the memory bandwidth of the full machine can be accessed by the one process. In addition, the \emph{multi} load layout will run the same process on each of the 16 cores, which simulates a use case where the full node is used for computation and every core uses the memory bandwidth of the socket. Since the performance results for the single and multi case do not differ qualitatively, we will only show the results for the multi case. 
This test setup is evaluated with four different configurations:
\begin{itemize}
  \item \emph{Real}: Baseline for performance comparisons where no complex numbers types are used.
  \item \emph{Complex}: Introduction of complex numbers, but no special handling is performed. This is the baseline for the complex number comparisons.
  \item \emph{Real handled}: The same as the \emph{Real} case, but with the new CoDiPack version that supports complex numbers.
  \item \emph{Complex handled}:  The same as the \emph{Complex} case, but with the new CoDiPack version that supports complex numbers.
\end{itemize}
These four configurations are checked against the four major CoDiPack types which are \emph{Linear Jacobian} (\ic{codi::RealReverse}), \emph{Reuse Jacobian} (\ic[breaklines]{codi::Real ReverseIndex}), \emph{Linear Primal} (\ic{codi::RealReversePrimal}) and \emph{Reuse Primal} (\ic{codi::RealReversePrimalIndex}).

\subsubsection{Memory comparison}
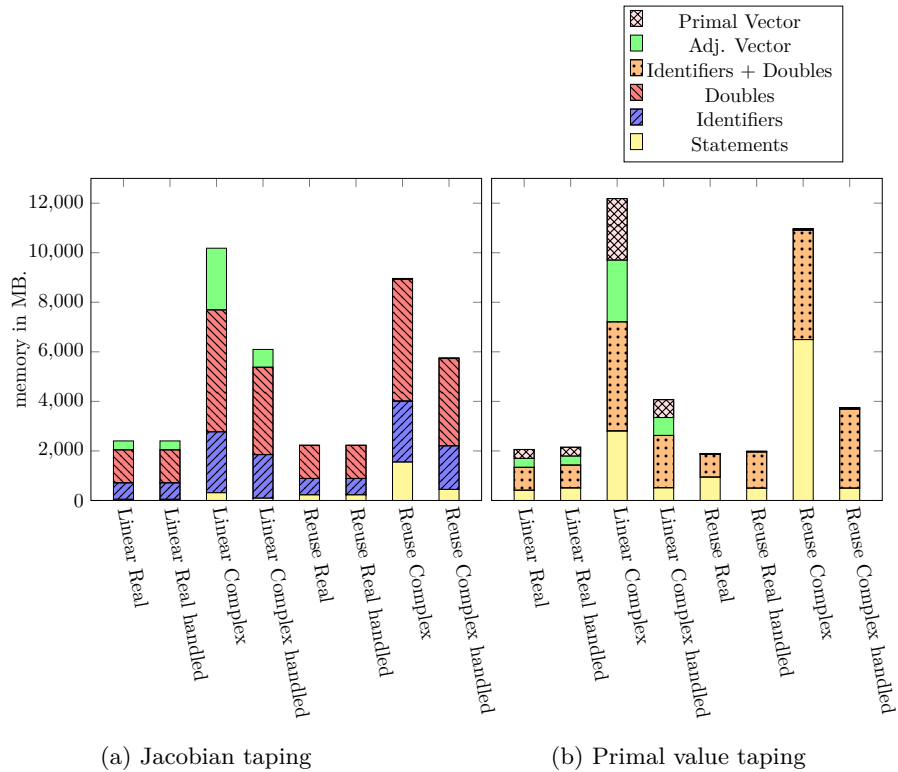
\begin{figure}
  \begin{subfigure}{0.45\textwidth}
    \scalebox{0.75}{
      \begin{tikzpicture}
        \begin{axis}[
          ybar stacked,
          ymin=0,
          ymax=13000,
          ylabel=memory in MB.,
          scaled y ticks = false,
          xtick=data,
          legend style={cells={anchor=west}, legend pos=north},
          reverse legend=true,
          xticklabels from table={\dataMemJacobian}{Label},
          x tick label style={rotate=-80,anchor=west, yshift=-0.0em},
        ]
          \memPlot{Stmts}{Statements}{black,fill=yellow!50}{\dataMemJacobian}
          \memPlot{Identifiers}{Identifiers}{black,fill=blue!50,postaction={pattern=north east lines}}{\dataMemJacobian}
          \memPlot{Doubles}{Doubles}{black,fill=red!50, postaction={pattern=north west lines}}{\dataMemJacobian}
          \memPlot{AdjVec}{Adj. Vector}{black,fill=green!50}{\dataMemJacobian}
          \legend{}
          \end{axis}
      \end{tikzpicture}
    }
    \caption{Jacobian taping}
  \end{subfigure}
  \hspace{1.5em}
  \begin{subfigure}{0.45\textwidth}
    \scalebox{0.75}{
      \begin{tikzpicture}
        \begin{axis}[
          ybar stacked,
          ymin=0,
          ymax=13000,
          ylabel=,
          yticklabel={\ },
          scaled y ticks = false,
          xtick=data,
          legend style={at={(0.9,1.05)},anchor=south east},
          reverse legend=true,
          xticklabels from table={\dataMemPrimal}{Label},
          x tick label style={rotate=-80,anchor=west, yshift=-0.0em},
        ]
          \memPlot{Stmts}{Statements}{black,fill=yellow!50}{\dataMemPrimal}
          \memPlot{Identifiers}{Identifiers}{black,fill=blue!50,postaction={pattern=north east lines}}{\dataMemPrimal}
          \memPlot{Doubles}{Doubles}{black,fill=red!50, postaction={pattern=north west lines}}{\dataMemPrimal}
          \memPlot{Identifiers_Doubles}{Identifiers + Doubles}{black,fill=orange!50,postaction={pattern=dots}}{\dataMemPrimal}
          \memPlot{AdjVec}{Adj. Vector}{black,fill=green!50}{\dataMemPrimal}
          \memPlot{PrimalVec}{Primal Vector}{black,fill=pink!50, postaction={pattern=crosshatch}}{\dataMemPrimal}
          
          \end{axis}
      \end{tikzpicture}
    }
    \caption{Primal value taping}
    \label{fig.memoryPrimal}
  \end{subfigure}
  \caption{Memory consumption for the real and complex numbers in different tapes.}
  \label{fig.memory}
\end{figure}

The memory comparison in Figure \ref{fig.memory} shows the data for the Jacobian taping approach on the left-hand side and the values for the primal value taping approach on the right-hand side.
For the Jacobian taping approach, the memory does not change for the new CoDiPack version (e.g., Linear Real vs. Linear Real handled). There is a slight memory increase for the primal taping approach due to the additional data size per statement. The memory increase is about 5\% (92 MB for Linear Real vs. Linear Real handled).

A large memory increase can be seen when switching from real numbers to complex numbers (e.g., Linear Real vs. Linear Complex). Adding the complex numbers to the expression framework of CoDiPack reduces the memory consumption of the tape. The intermediate values are no longer recorded, which leads to this improvement. The Jacobian approach reduces the memory by about $35$\% and the primal value approach by about $65$\%.

The reduction of the general memory consumption for complex numbers is significant. But it is also important to determine if the general memory increase from the real valued case to the complex case is optimal. The switch to complex numbers doubles the number of statements, and for each statement the number of arguments is doubled. Therefore, for Jacobian tapes, which store a statement for each component of the complex number, we expect the memory in the range from $2$ to $4$. For the Burgers case the memory factor is $2.6$ which is in the expected range. Since primal values tapes store only one statement for a complex assignment, the expected memory factor is in the range from $1$ to $2$ --- only the number of arguments is doubled. With a memory factor of $1.9$, the value for the Burgers case is in the expected range. 

The reduction of $35\%$ memory for the Jacobian taping approach and $65\%$ for the primal value taping approach show how important it can be to add complex numbers or other types to the expression template framework of AD tools.

\pgfplotstableread[col sep=comma]{
Label, Total, AdjVec, Stmts, Identifiers, Doubles
Linear Real, 2396.16, 361.38, 45.17, 665.60, 1331.20
Linear Real handled, 2396.16, 361.38, 45.17, 665.60, 1331.20
Linear Complex, 10188.80, 2488.32, 310.71, 2461.01, 4922.02
Linear Complex handled, 6082.56, 722.76, 90.34, 1761.28, 3522.56
Reuse Real, 2242.56, 16.50, 222.42, 665.60, 1331.20
Reuse Real handled, 2242.56, 16.50, 222.42, 665.60, 1331.20
Reuse Complex, 8990.72, 33.00, 1546.24, 2461.01, 4922.03
Reuse Complex handled, 5765.12, 33.00, 444.83, 1761.28, 3522.56
}\dataMemJacobian
\pgfplotstableread[col sep=comma]{
Label, Total, AdjVec, PrimalVec, Stmts, Identifiers_Doubles, Identifiers, Doubles
Linear Real, 2058.24, 361.38, 361.38, 406.55, 929.09, 0.0, 0.0
Linear Real handled, 2150.40, 361.38, 361.38, 496.89, 928.95, 0.0, 0.0
Linear Complex, 12185.60, 2488.32, 2488.32, 2795.52, 4413.44, 0.0, 0.0
Linear Complex handled, 4075.52, 722.76, 722.76, 504.47, 2119.68, 0.0, 0.0
Reuse Real, 1904.64, 16.50, 16.50, 934.15, 929.09, 0.0, 0.0
Reuse Real handled, 1996.80, 16.50, 16.50, 489.32, 1464.30, 0.0, 0.0
Reuse Complex, 10997.76, 33.00, 33.00, 6492.16, 4413.44, 0.0, 0.0
Reuse Complex handled, 3758.08, 33.00, 33.00, 489.32, 3194.88, 0.0, 0.0
}\dataMemPrimal

\pgfplotsset{symbolDefaults/.style={
  only marks,
  mark size=4pt}}

\newcommand{\timePlot}[4]{
  \addplot+[symbolDefaults] table[x index=#1, y expr=\thisrowno{0}+#2] {#4};
  \addlegendentry{#3}
}

\newcommand{\timePlotBelow}[4]{
  \addplot+[symbolDefaults, nodes near coords style={yshift=-12pt}, nodes near coords align=north] table[x index=#1, y expr=\thisrowno{0}+#2] {#4};
  \addlegendentry{#3}
}

\subsubsection{Time comparison}
\begin{figure}
  \begin{subfigure}[b]{0.55\textwidth}
      \begin{tikzpicture}
        \begin{axis}[
          height=4cm,
          width=0.9\textwidth,
          xlabel={time in sec.},
          ytick={-1, -2, -3, -4},
          ymax=-0.4,
          ymin=-4.6,
          xmin=0.1,
          xmax=0.6,
          yticklabels={Linear Jacobian, Reuse Jacobian, Linear Primal, Reuse Primal},
          legend columns=2,
          transpose legend,
          legend style={at={(0.76,1.01)},anchor=south west,/tikz/column 2/.style={column sep=5pt}},
          nodes near coords,
          nodes near coords style={yshift=3pt, color=black, scale=0.5},
          point meta=rawx
        ]
  
          \timePlot{1}{0.0}{Real}{results_new/epyc_gcc_real/codi_base_16.dat}
          \pgfplotsset{cycle list shift=1}
          \timePlotBelow{1}{0.0}{Real handled}{results_new/epyc_gcc_real/codi_complex_array_fix_16.dat}
          \legend{}
        \end{axis}
      \end{tikzpicture}
      \caption{Recording time multi}
    \end{subfigure}
    \hspace{0.25em}
    \begin{subfigure}[b]{0.45\textwidth}
      \begin{tikzpicture}
        \begin{axis}[
          height=4cm,
          width=\textwidth,
          xlabel={time in sec.},
          ytick={-1, -2, -3, -4},
          ymax=-0.4,
          ymin=-4.6,
          xmin=0.1,
          xmax=0.6,
          yticklabels={,,,},
          legend style={at={(1.,1.)},anchor=south east},
          nodes near coords,
          nodes near coords style={yshift=3pt, color=black, scale=0.5},
          point meta=rawx  
        ]
    
          \timePlot{4}{0.0}{Real}{results_new/epyc_gcc_real/codi_base_16.dat}
          \pgfplotsset{cycle list shift=1}
          \timePlotBelow{4}{0.0}{Real handled}{results_new/epyc_gcc_real/codi_complex_array_fix_16.dat}
%          \legend{}
        \end{axis}
      \end{tikzpicture}
      \caption{Reversal time multi}
    \end{subfigure}
\caption{Timings for the aggregated type handling in different tapes.}
  \label{fig.timeReal}
\end{figure}
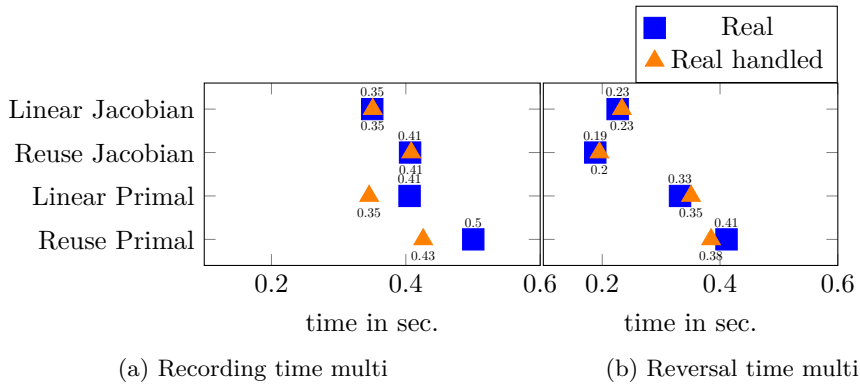

First, we want to compare the timing results of the new and old implementations. These are shown in Figure \ref{fig.timeReal}. Since the implementation of the Jacobian taping approach did not change, the timing results are nearly the same. For the primal value taping approach, we see an improvement in the results. The recording time is improved by at least 16\%. The reversal time increases for the primal value linear approach by $6$\% and is reduced for the primal value linear approach by $7$\%. The improvement in the recording is archived by reducing the number of function calls to the stack implementations. The changed times for the reversal come from data and implementation overhaul. They are now closer for both tape variants.

\begin{figure}
  \begin{subfigure}[b]{\textwidth}
    \begin{tikzpicture}
      \begin{axis}[
        height=5cm,
        width=0.9\textwidth,
        xlabel={time in sec.},
        ytick={-1, -2, -3, -4},
        ymax=-0.4,
        ymin=-4.6,
        xmax=6.4,
        yticklabels={Linear Jacobian, Reuse Jacobian, Linear Primal, Reuse Primal},
        legend columns=2,
        transpose legend,
        legend style={at={(1.0,1.01)},anchor=south east,/tikz/column 2/.style={column sep=5pt}},
        nodes near coords,
        nodes near coords style={yshift=3pt, color=black, scale=0.5},
        point meta=rawx
      ]
  
        \timePlot{1}{0.0}{Real}{results_new/epyc_gcc_real/codi_base_16.dat}
        \timePlot{1}{0.0}{Complex}{results_new/epyc_gcc_complex/codi_base_16.dat}
        \timePlotBelow{1}{0.0}{Real handled}{results_new/epyc_gcc_real/codi_complex_array_fix_16.dat}
        \timePlotBelow{1}{0.0}{Complex handled}{results_new/epyc_gcc_complex/codi_complex_tuple_fix_16.dat}
      \end{axis}
    \end{tikzpicture}
    \caption{Recording time}
    \vspace{2.em}
  \end{subfigure}
  
  \begin{subfigure}[b]{\textwidth}
    \begin{tikzpicture}
      \begin{axis}[
        height=5cm,
        width=0.9\textwidth,
        xlabel={time in sec.},
        ytick={-1, -2, -3, -4},
        ymax=-0.4,
        ymin=-4.6,
        xmax=6.4,
        yticklabels={Linear Jacobian, Reuse Jacobian, Linear Primal, Reuse Primal},
        legend style={at={(1.,1.)},anchor=south east},
        nodes near coords,
        nodes near coords style={yshift=3pt, color=black, scale=0.5},
        point meta=rawx  
      ]
  
        \timePlot{4}{0.0}{Real}{results_new/epyc_gcc_real/codi_base_16.dat}
        \timePlot{4}{0.0}{Complex}{results_new/epyc_gcc_complex/codi_base_16.dat}
        \timePlotBelow{4}{0.0}{Real handled}{results_new/epyc_gcc_real/codi_complex_array_fix_16.dat}
        \timePlotBelow{4}{0.0}{Complex handled}{results_new/epyc_gcc_complex/codi_complex_tuple_fix_16.dat}        
        \legend{}
      \end{axis}
    \end{tikzpicture}
    \caption{Reversal time}
  \end{subfigure}
\caption{Timings for the complex number handling in different tapes for the multi configuration.}
  \label{fig.timeComplexMulti}
\end{figure}
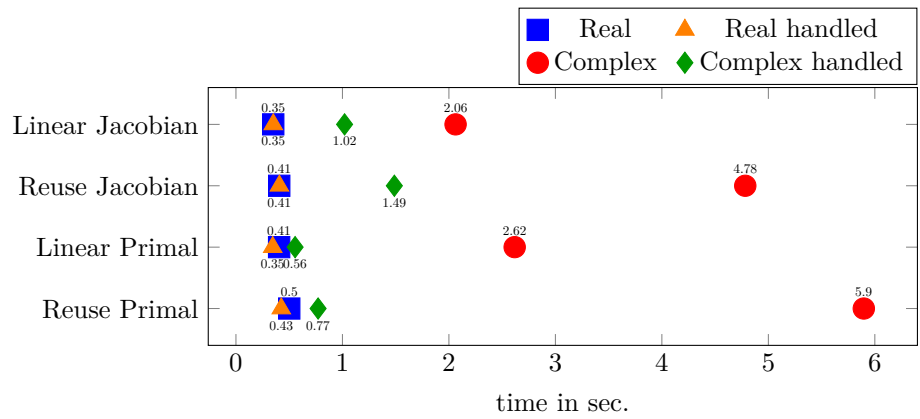
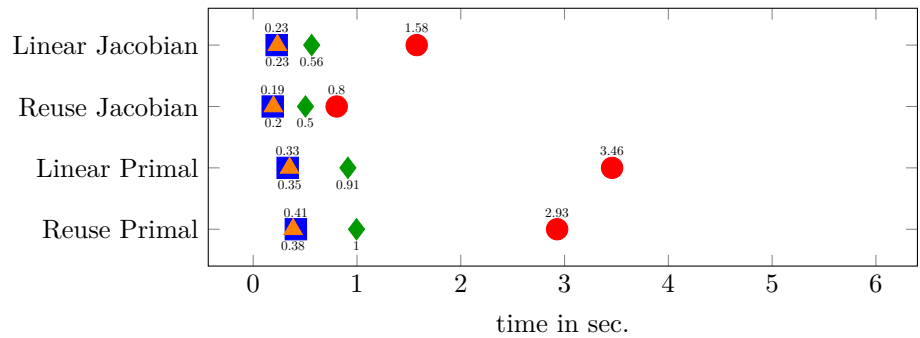

The results for the complex numbers are presented in Figure \ref{fig.timeComplexMulti}. It also contains the values for the real numbers as a reference. If the complex numbers are not handled, then the recording as well as the reversal time increases by a large factor. The additional overhead for the reuse index management during the recording comes from the index handling for the intermediate values. When the complex numbers are added to the expression templates of the AD tool, a large time improvement can be seen. The recording overhead for the linear Jacobian taping approach is $2.9$, which is close to the memory factor of $2.6$. In the reuse case, the timing factor is a bit larger due to the additional index handling, since the indices of the left-hand side can no longer be reused. The factor for the reversal time is around $2.5$ for both Jacobian taping approaches. All factors are within the expected range of $2$ to $4$ as discussed in the memory analysis.

The recording factor for the primal value taping approach is around $1.8$, which is also in the expected range from $1$ to $2$. The factors for the reversal are around $2.6$, which is higher than expected. An analysis of the assembly code did not yield any insight why this is the case.

\section{Conclusion}

We demonstrated how complex numbers can be integrated into an operator overloading AD tool that uses expression templates. The validity of the proposed solution is demonstrated with an implementation in CoDiPack. The performance values for a generic test case allow two conclusions. First, the overall performance of CoDiPack is not impacted by the additional handling of complex numbers. For primal value taping, an increased speed for the recording could be achieved due to the change of the data layout.
Second, when complex numbers are handled by the AD tool, the memory consumption, the tape recording speed and the tape reverse evaluation speed can be greatly improved for code parts that use complex numbers. The complex number handling is available in the $3.0$ version of CoDiPack.

\section*{Acknowledgement}
The authors gratefully acknowledge the funding of the German National High-Performance Computing (NHR) association for the Center NHR South-West.

\bibliographystyle{alpha}
\bibliography{literature}

\end{document}